\documentclass[twocolumn,pra,amsmath,aps,groupedaddress,superscriptaddress,floatfix,showpacs]{revtex4}
\usepackage{epsfig}
\usepackage{amsmath}
\usepackage{graphics}
\newcommand{\be}{\begin{equation}}
\newcommand{\ee}{\end{equation}}
\newcommand{\bea}{\begin{eqnarray}}
\newcommand{\eea}{\end{eqnarray}}
\newcommand{\ket}[1]{\left|#1\right\rangle}
\newcommand{\bra}[1]{\left\langle #1\right|}

\newcommand{\bc}{\begin{center}}
\newcommand{\ec}{\end{center}}

\renewcommand{\(}{\left(}
\renewcommand{\)}{\right)}
\renewcommand{\[}{\left[}
\renewcommand{\]}{\right]}
\newcommand{\forget}[1]{}

\newcommand{\re}{{\rm e}}
\newcommand{\ri}{{\rm i}}
\begin{document}
\title{Cavity-mediated  long-range interaction for fast multiqubit quantum logic operations}
\author{Kishore T.  Kapale}
\email{Kishor.T.Kapale@jpl.nasa.gov}
\affiliation{Quantum Computing Technologies Group, Jet Propulsion Laboratory, California Institute of Technology,
Mail Stop 126-347, 4800 Oak Grove Drive, Pasadena, California 91109-8099}
\author{Girish S. Agarwal\footnote{On leave of absence from Physical Research Laboratory, Navrangpura, Ahmedabad, 380 009, India}}
\affiliation{Dept.~of Physics, Oklahoma State University, Stillwater, Oklahoma 74078-3072} 
\author{Marlan O. Scully}
\affiliation{Inst.~for Quantum Studies \& Dept. of Physics, Texas A\&M University, College Station, Texas 77843-4242}
\affiliation{Dept.~of Chemistry, Princeton University, Princeton, NJ 08544}
\date{\today}
\begin{abstract}
Interactions among qubits are essential for performing two-qubit quantum logic operations. However, nature gives us only nearest neighbor interactions in simple and controllable settings. Here we propose a strategy to induce interactions among two atomic entities that are not necessarily neighbors of each other through their common coupling with a cavity field. This facilitates fast multiqubit quantum logic operations through a set of two-qubit operations. Using its explicit position dependence, this interaction can be employed for simulation of quantum spin systems. The ideas presented here are applicable to various quantum information proposals for atom based qubits such as, trapped ions, atoms trapped in optical cavities and optical lattices.
\end{abstract}
\pacs{03.67.Lx, 42.50.pq, 32.80.Qk, 03.67.-a}
\maketitle

\section{introduction}
Quantum information science~\cite{Nielsen:2000} has made rapid progress recently with scalable architectures proposed for atom based qubits through the ion-trap schemes~\cite{iontraplogic} and for photonic qubits through the linear optical quantum computing~\cite{Knill:2001} and cavity quantum electrodynamics (QED)~\cite{Duan:2004} schemes, along with various other proposals for quantum logic operations for atomic~\cite{Gabris:2004,Goto:2004,atomiclogic},  photonic~\cite{Zubairy:2003} and hybrid~\cite{Tian:2004a} qubits. 

Two-qubit quantum logic operations require interaction between the physical entities used for encoding the qubits.  Physical systems provide natural grounds for implementations of two-qubit operations as there are plenty of cases providing controllable interaction between two entities. However, direct multiqubit operations require controllable interactions between more than two entities at a time that are difficult to come by. Thus, the decomposition of multiqubit  gates into their two-qubit and single-qubit counterparts is an essential step in quantum circuit design and implementations. Moreover, majority of the two-body interactions have a spatially dependent interaction strength. Thus, in most cases only near-neighbor interactions are available.

To this end, we note that it is possible  to  induce interaction between atoms coupled collectively to the cavity vacuum~\cite{Agarwal:1997}. This interaction has been shown to be useful to perform quantum logic operations by G\'{a}bris and Agarwal~\cite{Gabris:2004} in the case of two level atoms trapped inside a cavity.  Here we  recognize that the interaction induced between the atoms through their common coupling with the cavity vacuum is essentially independent of the distance between the atoms.  We employ the long-range nature of this interaction and develop a scheme to allow any two atoms from a chain of trapped neutral atoms  to interact with each other. This interaction is then exploited to perform quantum logic operations between any two qubits. Our model employs metastable atomic states as qubits, thus, qubit decoherence is not an issue. Moreover, the atom-cavity coupling is dispersive in nature;  therefore, cavity does not contain any real photons at any stage of the interaction. Thus, the cavity decay becomes a non-issue as well.  It is well known that multiqubit quantum logic operations can be achieved through  a sequence of single qubit and two-qubit gates (e.g., controlled-NOT (CNOT) and controlled-sign (C-Sign) gates)~\cite{Barenco:1995a}. Thus,  our scheme can be easily extended to perform multiqubit operations. In fact, as we show later, the quantum circuits for the multiqubit quantum operations as they are usually drawn, which involves several non-neighbor qubit operations,  can be directly implemented through of a sequence of operations using our non-local scheme.

It is also instructive to recollect that an important hurdle for scalability has been identified for the ion-trap quantum computing proposals~\cite{iontraplogic}. Namely,  physical motion of ions is required to ensure proximity among the qubits for the two-qubit logic operation. This dictates tremendous speed constraints on the current ion-trap quantum computer architecture. Very recently Tian {\it et al.}~\cite{Tian:2004a} have proposed a hybrid qubit approach, by coupling the ion-trap qubits with the superconducting ones,  to cure this pathology of trapped-ion systems. Nevertheless, introduction of non-neighbor interactions would provide tremendous speed-up for the trapped ion proposals. Furthermore, devising simple strategies for implementation of the error correcting codes~\cite{DiVincenzo:1996} is essential, especially, in the context that  the experimental demonstration has been possible only for a three-bit code~\cite{Chiaverini:2004} so far. 

The article is organized as follows. As we are proposing a non-neighbor interaction for quantum logic operations we present  a simple analysis of how many operations are required in the conventional setting to carry out a non-local CNOT gate between the first and the $N$th qubit from a set of total  $N$ qubits. Next, we present how distance-independent interaction can be induced between any two qubits from a chain having  a number of them. We give a set of operations required to perform the quantum phase gate operation such that the non-local nature is maintained. Then, we clarify advantages offered by our scheme and contrast it with several other quantum logic schemes for the ionic and neutral atom qubits. Finally we present our conclusions. In the appendices we provide  calculational details leading to the quantum logic operations.

\section{Advantages of The Non-neighbor Interactions}
\begin{figure*}
\includegraphics[width=0.85\textwidth]{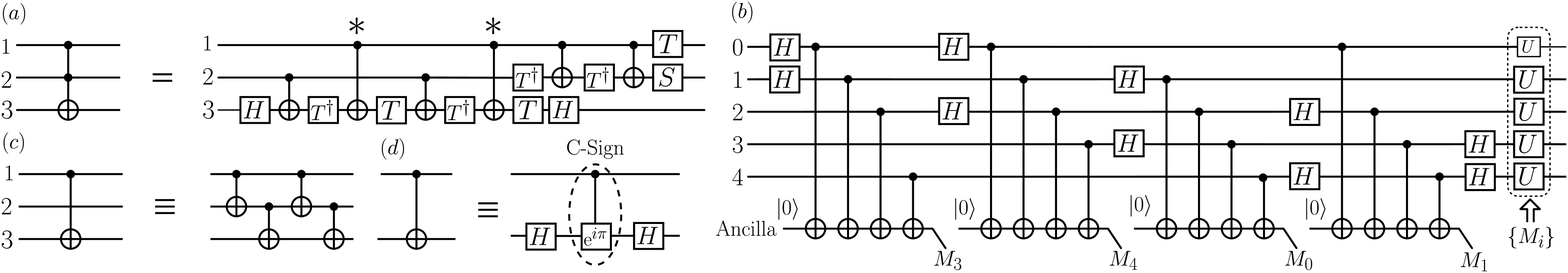}
\caption{\label{Fig:Circuits} Illustration of the need of non-neighbor interactions for fast quantum logic operations. $(a)$ Decomposition of a three-qubit Toffoli gate into two qubit CNOT gates and single qubit unitary transformations (See Eq.~\eqref{Eq:SQUM}) adapted from Ref.~\cite{Nielsen:2000}.  The steps denoted by $*$ involve non-neighbor interaction. $(b)$ An error correcting network for a five bit encoding~\cite{DiVincenzo:1996}. Error syndrome measurements ($\{M_3, M_4, M_0, M_1\}$) on the ancillas dictate the corrective unitaries ($U$) to be performed in the end to protect the encoded qubits from various errors. Notice the need for several non-neighbor two-qubit CNOT operations. $(c)$ Decomposition of a Non-neighbor two-qubit CNOT gate into near-neighbor two qubit gates. $(d)$ Equivalence between the CNOT and C-Sign gates modulo single qubit unitary transformations.}
\end{figure*}

To emphasize the speed-up obtainable through non-neighbor interactions we revisit the design of multiqubit quantum logic gates through a sequence of single qubit 
unitary transformations and two-qubit operations. For example, we consider the circuit representation of the three-qubit Toffoli gate in Fig.~\ref{Fig:Circuits}$(a)$ in terms of the CNOT gates and single qubit unitary transformations,
\begin{align}
\label{Eq:SQUM}
H =\frac{1}{\sqrt{2}} \(\begin{matrix} 1 &  1 \\
                               1 & -1
  \end{matrix}\)\!,\,
S=\(\begin{matrix} 1 &  0 \\
                   0 & \ri
  \end{matrix}\)\!, \text{ and }\,
T = \(\begin{matrix} 1 &  0 \\
                   0 & \re^{\ri \pi/4}
\end{matrix}\)\!.
\end{align}
It needs at least two non-neighbor CNOT gates.
In Fig.~\ref{Fig:Circuits}$(b)$ we consider a five-bit error correction network\forget{as discussed in detail by Divincenzo and Shor}~\cite{DiVincenzo:1996}. It needs multiple interactions of each encoding qubit with the ancilla, therefore with only near-neighbor interactions there would be a dramatic increase in the number of operations needed for its implementation. 

To quantify the number of  extra operations required per non-neighbor two-qubit gate we
consider an example in Fig.~\ref{Fig:Circuits}$(c)$ showing a decomposition of 
non-neighbor CNOT operations through multiple near-neighbor CNOT operations. Thus, to perform a non-neighbor two qubit gate between the first and third qubit from a set of total three qubits the optimal sequence of operations (as given in Fig.~\ref{Fig:Circuits}$(c)$) requires three more operations. It turns out,  however, that this strategy is not optimal if it is extended in a straightforward manner to a set of total $N$ qubits for performing two qubit operation between the first and the $N$th qubit. The optimal strategy, using only local or near-neighbor interactions,  is obtained by swapping the $N$ the qubit with the `$N-1$' th qubit, then swapping the `$N-1$' th qubit with the `$N-2$' th one and so on till the state of the $N$th qubit is transferred to the $2$nd qubit.A simple calculation shows that this swapping operation requires $N-2$ operations. Then the two qubit CNOT could be performed between the near-neighbor first and second qubits. Once again $N-2$ SWAP operations would be needed to bring back the new state of the second qubit to the $N$th qubit where it belongs. It is also known that a single SWAP operation requires three CNOT operations. Thus, to perform a two-qubit non-local gate between the first and the $N$th qubit one requires $6(N-2)$ extra operations. 

It has to be borne in mind, however, that each SWAP operation has to be completely error-free otherwise the above procedure would introduce tremendous amount of uncorrectable errors. Thus, one needs each CNOT operation implemented in a fault-tolerant manner, which would require, at the least say, ten error-correcting operations per CNOT gate. Thus, actual number of operations require would be  $60(N-2)$ operations. To illustrate,  for a moderate number of qubits, say ten, total 480 extra operations would be needed just to perform a single two-qubit gate between the first and the tenth qubit. 

In the above calculation we have assumed that the error correcting network needed to perform a single CNOT in a fault-tolerant manner does not require any non-local operations. However, it can be easily seen from Fig.~\ref{Fig:Circuits}$(b)$ that this is not the case. Thus, the actual number of operations would be much higher than $60(N-2)$.

Therefore, as the complexity of the quantum circuit increases, more and more non-neighbor interactions would be  needed and schemes based on conventional approach would  be very slow at best and very error-prone at worst. Moreover, the number of  operations needed to perform a single non-neighbor gate between the first and the last qubit would increase with the system size in the conventional setting. Therefore, the advantage offered by efficient non-local interaction can not be overemphasized.

Just to note, the single qubit operations of Eq.~\eqref{Eq:SQUM} can be attained through properly timed Raman pulses coupling the two qubit-levels $\ket{0}$ and $\ket{1}$; this is routinely done in cavity-QED systems. Therefore, we restrict ourselves to the non-neighbor quantum phase gate operation. 
\forget{do not delve into their implementations of the single qubit unitary transformations and point out that they are fairly straightforward to obtain for the cavity-QED systems.}

\section{Non-local interaction through cavity vacuum}
In the discussion to follow we describe our scheme in detail and show how distance- independent interaction can be introduced between a pair of atoms through their common interaction with the cavity vacuum.

To accomplish scalable architecture it should be possible  to perform two-qubit operations with equal ease in the presence of other qubits. To illustrate, having only two-level systems as qubits interacting dispersively with the cavity field as considered by G\'{a}bris {\it et al.}~\cite{Gabris:2004}, one needs twice as many operations to accomplish two qubit gate in the presence of an extra atom in the cavity as this third atom also takes part in the collective interaction.  Therefore we employ collective coupling with the cavity with the choice of turning on the interaction as needed instead of having it on at all times. 

\begin{figure}
\includegraphics[scale=0.22]{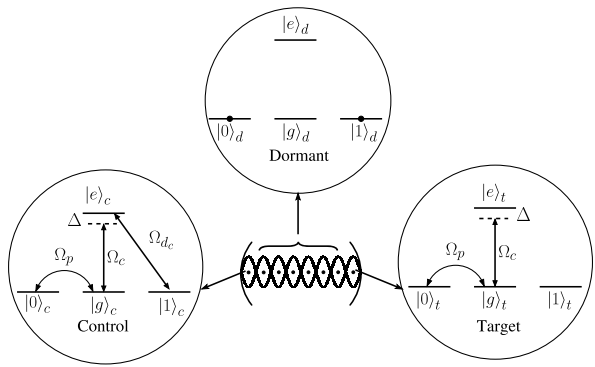}
\caption{\label{Fig:Scheme} Scheme for the non-neighbor two-qubit quantum phase gate with cavity mediated interaction. The cavity mode is coupled to the $\ket{g}-\ket{e}$ transition for all the atoms with detuning $\Delta$.  The external drive fields couple different atomic states as per the Hamiltonian in Eq.~\eqref{Eq:ExtHamiltonian}. The external coupling could be direct or through Raman pulses as discussed in the context of Eq.~\eqref{Eq:ExtHamiltonian}. The dormant atoms maintain their general state $\alpha\ket{0}_d +\beta\ket{1}_d$ while quantum logic operation is being performed on the control and target atoms. The atomic energy levels $\ket{0}, \ket{g}$ and $\ket{1}$ are shown to be degenerate just for convenience. This is not a requirement for the success of the proposal. 
}
\end{figure}

We consider a linear array of $N$ atoms placed in a cavity. We note that the cavity supports standing wave field with spatially dependent field amplitude. Therefore, it is important to trap atoms such that they all see the same field strength. Such an architecture can be achieved using the proposals for trapping atoms inside optical cavities~\cite{Pinkse:2000}, through the marriage of ion trap and cavity-QED systems~\cite{cavitytrapion} or in the chain of neutral atoms trapped in standing wave fields~\cite{Schrader:2004}. The atoms are assumed to be identical and 
have a four-level internal structure as shown in Fig.~\ref{Fig:Scheme}. 
The states forming qubits could be taken as hyperfine sublevels of an electronic states or two states of  a single hyperfine manifold. The preparation of the qubit state can be accomplished through efficient mixing of the well developed optical pumping and adiabatic population transfer techniques. The states representing the qubit are 
chosen so that they do not directly interact with the cavity field. However, each atom has 
an extra pair of levels that can interact with the cavity field. Thus, once the $i$th, and $j$th atoms are brought into these levels they interact with the cavity field through the
Hamiltonian
\begin{equation}
\label{Eq:Hamiltonian}
{\mathcal H} =  {\mathcal H}_0 -\hbar \Omega_c \,\sum_{l=i,j}(\ket{e_l}\bra{g_l} a_k + \ket{g_l}\bra{e_l} a_k^\dagger ) \,,
\end{equation}
with $\ket{e_i}$ and $\ket{g_i}$ being the levels of atom $i$ close to resonance with the cavity field
Also, $a_k$ is the cavity mode annihilation operator and $\Omega_c$ is the coupling strength of the atomic transition with the cavity field.
Here the free Hamiltonian is 
\begin{equation}
{\mathcal H}_0 = \hbar \nu_k \, a_k^{\dagger} a_k + \hbar \omega_{g_i} \ket{g_i}\bra{g_i} + \hbar \omega_{e_i} \ket{e_i}\bra{e_i}\,,
\end{equation}
including only the relevant atomic energy levels. The atomic energies ($\hbar \omega_{g_i}, \hbar \omega_{e_i}$) are measured with respect to the ground state, $\ket{0_i}$, of the atoms. Here the position dependence of the Rabi frequency and the atomic dipole operators is not shown as the atomic positions (for example, anti-nodes of the standing wave field of the cavity)  are chosen such that all atoms see the same cavity field strength. By including the position dependence of the cavity field and the atomic dipoles, this Hamiltonian ~\eqref{Eq:Hamiltonian} becomes tunable and can be employed for simulation of the quantum spin systems~\cite{Porras:2004}. It is important to note that the cavity field is not directly resonant with the $\ket{e}-\ket{g}$ transitions and it is coupled only dispersively. Therefore, the cavity field is  in its vacuum state and the atom does not get excited by the cavity field or if it is in the excited state it does not emit photon in the cavity mode. Thus, through adiabatic elimination of the states corresponding to the presence of photons in the cavity mode we arrive at the effective interaction Hamiltonian~\cite{Agarwal:1997},
\begin{align}
\label{Eq:EffHamiltonian}
{\mathcal H}_{\rm eff} &= \hbar \eta\, \Bigl(\sum_{k=i,j}\!\ket{e_k}\bra{e_k} \nonumber \\
&+ \ket{e_i}\bra{g_i}\otimes\ket{g_j}\bra{e_j}
+ \ket{e_j}\bra{g_j}\otimes\ket{g_i}\bra{e_i} \Bigr)
\end{align} 
where $\eta = \Omega_c^2/\Delta$, and $\Delta$ is the cavity field detuning with respect to the atomic transition. We have assumed $\Delta\gg \Omega_c$ for arriving at this result. The first term leads to trivial phase factors that can be corrected in a straightforward manner and the last two terms lead to coupling between the atom through virtual exchange of a cavity photon. Thus, the cavity vacuum effectively induces interaction between two atoms, immaterial of their spatial position provided they both see the same cavity field strength. The dipole-dipole interaction usually falls of as inverse sixth power of the distance between the dipoles. Whereas, the interaction induced between the atomic dipoles through the cavity vacuum is independent of the distance between them. This long-range interaction  can be employed to perform non-local quantum logic operations.

\section{The non-local quantum phase gate}
In this section we provide the set of operations needed to accomplish direct two-qubit operation between any of the two qubits from a linear array of $N$-qubits. We note that, the two qubit CNOT gate can be decomposed into Hadamard transformations (See Fig.~\ref{Fig:Circuits}$(d)$, and matrix $H$ from Eq.~\eqref{Eq:SQUM}) on the target qubit and a C-Sign gate between the two qubits. Therefore, we only resort to implementing the C-Sign gate operation given by
\begin{equation}
\sum_{j,k=0,1} c_{jk} \ket{j,k} \longrightarrow \sum_{j,k=0,1} (\re^{\pm \ri\pi})^{j\, k} c_{jk} \ket{j,k}\,. 
\end{equation} 
The calculational details are given in Appendix B at length and the choice of the operations steps taken is justified.

Now we analyze several possible initial states of the atoms and their interaction with the cavity field. The results are summarized below: It can be seen that the states $\ket{g_i, g_j,0_k}$, $\ket{g_i, a_j, 0_k}$ and $\ket{a_i, g_j,0_k}$ remain unaffected by the cavity-field. Here state $a_{i,j}$ corresponds to some arbitrary atomic level $a$ that does not interact with the cavity field. Also if the interaction time is taken to be $\eta t = \pi$, the states $\ket{g_i, e_j, 0_k}$ and $\ket{e_i, g_j,0_k}$ return to their original atomic configurations, and the states 
$\ket{e_i, a_j, 0_k}$  and $\ket{a_i, e_j, 0_k}$ acquire a phase factor of $\re^{-\ri \pi}$.

Another important ingradient necessary for our proposal is selective addressing of the atoms. To achieve this we consider a general interaction Hamiltonian
\begin{equation}
\label{Eq:ExtHamiltonian}
{\mathcal H}_{\rm ext} = -\hbar\Bigl[ \Omega_p \re^{\ri \phi_p}\!\!\sum_{i=c,t}\! \ket{g}_i \bra{0}_i +  \Omega_{d_c} \re^{\ri \phi_{d_c}} \ket{e}_c \bra{1}_c + {\rm H. c.} \Bigr].
\end{equation}
describing application of external optical fields on certain atomic transitions.
Here the Rabi frequencies could be directing couplings between the involved levels or they could be composite Rabi frequencies of a couple of Raman pulses coupling the involved levels through intermediate levels $\ket{i_{1,2}}$. In the composite case the Rabi frequencies can be written as
\begin{equation}
\Omega_p = \frac{\Omega_{0,i_1} \Omega_{i_1,g}}{\delta_1},\quad
\Omega_{d_c} = \frac{\Omega_{1_c,i_2} \Omega_{i_2,e_c}}{\delta_2}\,.
\end{equation}
Here $\Omega_{j,k}$ denotes the Rabi frequency of interaction of  levels $j$ and $k$ with the corresponding light field applied with detuning $\delta_{1}$ or $\delta_{2}$ on the $j\text{-}k$ transition.
It can be noted that, the composite Raman pulses are routinely used in the ion-trap quantum logic gates. We note that, complete transfer of population, $\ket{0}\rightarrow\ket{g}$, can be accomplished through an application of a pulse with parameters $\Omega_p t = \pi/2$ and $\phi_p = 3 \pi/2$ and the inverse operation, $\ket{g}\rightarrow\ket{0}$, with $\Omega_p t = \pi/2$ and $\phi_p = \pi/2$. Similar considerstions hold for the pulse with Rabi frequency $\Omega_{d_c}$ resonant on the $\ket{1}_c-\ket{e}_c$ transition of the control atom. The details of why a specific phase of the Rabi frequency is necessary to achieve population transfer is discussed in Appendix A.

Using the characteristics of the atom-cavity interaction and
selective addressing through the external fields , as discussed in detail in Appendix B, 
we devise a set of operations for the C-Sign gate:
\begin{enumerate}
\item Operation $\ket{0}\rightarrow \ket{g}$ on both the target and control atoms through a pulse of Rabi frequency $\Omega_{p}\re^{\ri \phi_p}$ with $\Omega_{p}t = \pi/2$ and $\phi_p=3\pi/2$.
\item Operation $\ket{1}_c\rightarrow \ket{e}_c$ through a
pulse of Rabi frequency $\Omega_{d_c}\re^{\ri \phi_{d_c}}$  with $\Omega_{d_c} t = \pi/2$ and $\phi_{d_c}=3\pi/2$ to move the control qubit to the excited state interacting with the cavity.
\item Interaction of the control and target atoms with the cavity for the time $t = \pi/\eta$.
\item Operation $\ket{e}_c\rightarrow \ket{1}_c$, to bring back the qubit state of the control qubit, through the same pulse as in step 2 except for the phase $\phi_{d_c}=\pi/2$.
\item Operation $\ket{g}\rightarrow \ket{0}$ on both control and target via the same pulse of step 1, and the phase $\phi_p=\pi/2$.
\end{enumerate}
We note that, steps (1,5) and (2,4) can be accomplished via appropriate terms in Eq.~\eqref{Eq:ExtHamiltonian}. The effect of these operations on various initial states of the two-qubits is summarized at length in Table~\ref{Table}. By choosing the interaction time $t$ with the cavity vacuum mode such that $\eta t = \pi$, the desired two-qubit C-Sign gate operation is obtained. 

The fidelity calculation for the above model of the two qubit gate is summarized in Appendix C. Our numerical studies show a gate fidelity of 99\% , with the cavity decay ($\kappa$), the spontaneous decay of level $\ket{e}$ ($\gamma$) and the detuning ($\Delta$) taken  to be  $0.01\, \Omega_c$, $0.0001\, \Omega_c$ and $10\, \Omega_c$ respectively. For experimental cavity parameters~\cite{Pinkse:2000} ($\Omega_c=32 \pi\,  \text{MHz}$, and $\kappa = 2.8 \pi\,  \text{MHz}$), a modest $\gamma=0.001\,\Omega_c$ and $\Delta= 10\,\Omega_c$ we obtain a fidelity of 93\%. 
Once again we would like to point out that all the states $\ket{0}$, $\ket{1}$, $\ket{g}$ and $\ket{e}$ for all the atoms could be taken to be metastable  and the external coupling achieved through Eq.~\ref{Eq:ExtHamiltonian} could be achieved through Raman pulses. Therefore, atomic decoherence is not an issue. We have provided above fidelity calculations just for completeness. In such a case the atomic transition $\ket{g}\text{-}\ket{e}$ could be taken in the microwave range. The scheme could very well be applicable in the optical range, one only needs an appropriate metastable level $\ket{e}$ so that decoherence does not remain an issue. 

To come back to the gate operations, employing Hadamard transformation, $H$ from Eq.~\eqref{Eq:SQUM},  on the target qubit before and after the C-Sign gate,  one obtains a CNOT gate as shown in Fig.~\ref{Fig:Circuits}$(d)$. Combining several of these CNOT gates as shown in Fig.~\ref{Fig:Circuits}$(a)$ and $(b)$, the three-qubit Toffoli gate and five-bit error-correcting network can be directly constructed. 
\begin{table}[ht]
\caption{\label{Table}Two-qubit non-neighbor C-Sign gate operation: The effect of step 3 on the state $\ket{e_c, g_t}$ is non-trivial and it leads to the state $\re^{-\ri \eta t}[\cos(\eta t) \ket{e_c, g_t} - i  \sin(\eta t) \ket{g_c, e_t}]$. Thus,  with the choice of $\eta t = \pi$ it becomes
 as $\ket{e_c, g_t}$ as shown below. Moreover, the state $\ket{e_c, 1_t}$ acquires a phase factor $\re^{-\ri \eta t}= \re^{-\ri \pi}=-1$ under step 3. It can be noted that the first and the last column taken together correspond to the truth table of the C-Sign gate.}
 \vskip0.2cm
{\footnotesize
\begin{tabular}{|c|c|c|c|c|c|c|c|}
\hline 
input & step 1 & step 2 & step 3 & step 4 & step 5 & output  \\
\hline
$\ket{0_c,0_t}$ & $\ket{g_c, g_t}$ &  $\ket{g_c, g_t}$ &  $\ket{g_c, g_t}$ & $  \ket{g_c, g_t}$  & $ \ket{0_c, 0_t}$ & $\ket{0_c,0_t}$ \\
\hline
$\ket{0_c,1_t}$ & $\ket{g_c, 1_t}$ &  $ \ket{g_c, 1_t}$ &  $ \ket{g_c, 1_t}$ & $ \ket{g_c, 1_t}$ & $\ket{0_c, 1_t}$ & $\ket{0_c,1_t}$ \\
\hline
$\ket{1_c,0_t}$ & $\ket{1_c, g_t}$ &  $\ket{e_c, g_t}$ &  $ \ket{e_c, g_t}$ & $\ket{1_c, g_t}$ & $\ket{1_c, 0_t}$ & $\ket{1_c,0_t} $\\
\hline
\forget{ & & & + \sin(\eta t)\ket{g_c, e_t} & + \sin(\eta t)\ket{g_c, e_t}  & & & \\
\hline}
$\ket{1_c,1_t}$ & $\ket{1_c, 1_t}$ &  $\ket{e_c, 1_t}$ &  $-\ket{e_c, 1_t}$ & $-\ket{1_c, 1_t}$& $-\ket{1_c, 1_t}$ & $-\ket{1_c,1_t}$\\
\hline
\end{tabular}
}
\end{table}

The cavity-atom interaction~\eqref{Eq:Hamiltonian} also provides a single-step mechanism to create entanglement between distant atoms. For example, an initial state of two atoms $\ket{e_i, g_j}$ after the interaction with the cavity vacuum for the time $\eta t = \pi/4$ gives the entangled state 
\begin{equation}
\frac{1}{\sqrt{2}}\re^{-\ri \pi/4}(\ket{e_i, g_j} -i \ket{g_i, e_j})\,,
\end{equation}
which can be transformed into any of the four Bell-states by one-qubit unitary transformations.

\section{Advantages of the Current Proposal and its Connection with prior proposals}
In this section we mention advantages of our proposal and contrast it briefly with some of the other cavity QED and ion trap quantum phase gate proposals in the literature.

It is interesting to contrast the proposed scheme with the ones in the literature for atomic qubits.  Pellizzari {\it et al.}~\cite{Pellizzari:1995} have proposed a scheme for the implementation of controlled unitary gates through adiabatic passage on a  Raman transition. Their proposal hinges on transferring the qubit from two atoms to an extra pair of levels within a single atom and requires the atoms to be close to each other. The atomic level scheme for two qubit operations requires three doubly degenerate, i.e. in effect six, energy levels for each atom.  Another contrasting feature of this proposal is that it induces interaction among atomic qubits by the exchange of a real cavity photon, thus it is susceptible to the cavity decay.

Further, the proposal by Cirac and Zoller~\cite{Cirac:1995} for cold trapped ions achieves non-local operations through a collective excitation of the  vibrational motion of the ions with lasers. In the context of multiqubit operations, Goto and Ichimura~\cite{Goto:2004} propose a cavity QED based scheme to perform multiqubit unitary gate by adiabatic passage.  The gate operation mechanism is completely different and hinges on the presence of cavity photons, therefore it is susceptible to the cavity decay. 

Physically our scheme is close to the one considered by G\'{a}bris and Agarwal~\cite{Gabris:2004}, which uses two level atoms interacting with the cavity. The physical closeness comes in the sense that the qubit-qubit interaction is mediated by the cavity vacuum.  However, the sequence of operations proposed by them is very different and therefore require more number of operations for multiqubit gates as opposed to the possibility of direct circuit implementation available in the  present scheme. Another proposal using cavity mediated interaction is by Zheng and Guo~\cite{Zheng:2000};
 however, it does not employ the long-range nature of this interaction. Jan\'{e} et al.~\cite{Jane:2002} utilize the cavity field for quantum logic  with two three-level atoms and mention that the atoms could be arbitrarily positioned in the cavity field at integral wavelength separation, but do not consider simultaneous presence of more than two atoms in the cavity field. As it is already clear from the proposal in Ref.~\cite{Gabris:2004}, even though the interaction could be introduced by similar means, the actual set of operations performing the two-qubit gates are very crucial in determining if two-qubit gates can be performed in the presence of other qubits or not. The non-local interactions can be easily introduced if the presence of other qubits does not require alteration of the gate operation sequence.

Thus, compared to several prior proposals for multiqubit gate operations among the atomic qubits,  our scheme is not susceptible to the cavity decay as it only uses non-local coupling available through the interaction with the cavity vacuum. Moreover, the two qubit gate operations required do not depend on the presence of a large number of other qubits in the system. Therefore, the sequence of operations can be applied in succession to the qubits of interest allowing direct quantum circuit implementation of any multiqubit quantum logic operation and error correction networks. The atomic qubits are implemented through metastable atomic levels; therefore, qubit decoherence is a non-issue as well.

\section{Conclusion}

To summarize, we have demonstrated a non-neighbor two-body interaction between atomic qubits through their collective coupling with the cavity vacuum mode. This non-neighbor interaction can be employed to obtain implementations of the two-qubit universal quantum gates. Thus, we provide an architecture for performing fast quantum logic operation in the presence of other qubits. As selective coupling between any two qubits  becomes available, immaterial of their spatial position, multiqubit operations can be quickly performed through a sequential application of laser pulses to the appropriate atoms. Several advantages offered by our scheme include, practically no decoherence as only the metastable atomic states are used and the cavity is always in the vacuum state.  The proposal is fairly general and can be applied to variety of sytems using atomic qubits, such as ion-traps, trapped atoms or ions in optical lattices or cavities. Most importantly this approach provides a strategy, using current experimental techniques, to surpass the pathology of the ion-trap quantum computing proposals that require movement of the ions to facilitate two-body interactions. 

\acknowledgments
Part of this work was carried out (by K.T.K.) at the Jet Propulsion Laboratory under 
a contract with the National Aeronautics and Space Administration (NASA). KTK wishes to thank Prof.~J.~P. Dowling and Dr.~F.~Spedalieri for useful discussions and acknowledges support from the National Research Council and NASA, Codes Y and S.
G.S.A. acknowledges support from National Science Foundation, Grant No. NSF-CCF 0524673.  M.O.S. wishes to acknowledge support from Air Force Research Lab (AFRL, Grant No.  F30602-01-1-0594), Air Force Office of Scientific Research (AFOSR, Grant No. FA9550-04-1-0206), and TAMU Telecommunication and Informatics Task Force Initiative (TITF, Grant No. 2001-055).

\appendix

\section{Phase free population transfer}
In this appendix we show how to obtain population transfer of the type $\ket{b}\rightarrow \ket{a}$ with the usual $\pi$ pulses. Normally application of a $\pi$  pulse achieves population transfer but imparts an extra phase factor to the final state.  While carrying out quantum logic operations  these phase factors become relevant and it is better to avoid them.  Here we show, how the extra phase factors could be eliminated to obtain clean state transfer by appropriately phased optical pulses.  This technique is essential for performing several of the operations needed to obtain the quantum phase gate as described in the main body of the paper.

Consider interaction of a two level atom (lower level $\ket{b}$ and upper level $\ket{a}$ with an optical field of Rabi frequency $\Omega(\exp{\ri\,\phi})$ resonant on the transition. This can be described by the interaction Hamiltonian
\begin{equation}
H = - \hbar \Omega \bigl[\ket{a}\bra{b}\exp(\ri\, \phi) + \ket{b}\bra{a}\exp(-\ri\, \phi) \bigr]\,.
\end{equation}
Thus for a general wavefunction $\ket{\Psi}= a(t) \ket{a} + b(t) \ket{b}$ the population rate equations can be written through the Schr\"{o}dinger equation
\begin{equation}
\dot{\ket{\Psi}} = - \frac{\ri}{\hbar} H \ket{\Psi}\,,
\end{equation}
as
\begin{align}
\dot{a}(t) &= \ri\, \Omega \exp(\ri\, \phi) b(t) \\
\dot{b}(t) &= \ri\, \Omega \exp(- \ri\, \phi) a(t)\,.
\end{align}
The general solution of this set of coupled equations, in terms of the initial values $a(0)$ and $b(0)$, can be written as
\begin{align}
a(t) &= a(0)\cos(\Omega t) + \ri\, b(0) \exp(\ri\, \phi)\sin(\Omega t)\,, \\  
b(t) &= b(0)\cos(\Omega t) + \ri\, a(0) \exp(-\ri\, \phi) \sin(\Omega t)\,. 
\end{align}
Therefore, to obtain a clean state transfer of the kind $\ket{b}\rightarrow \ket{a}$ we initially have $a(0) = 0, b(0)=1$ and we need 
$\Omega t = \pi/2$ and $\ri\, \exp(\ri\, \phi)= 1$, i.e., $\phi=-\pi/2$ or $\phi = 3\pi/2$.
Also to obtain  $\ket{a}\rightarrow \ket{b}$ we have $a(0) = 1, b(0)=0$ and we need 
$\Omega t = \pi/2$ and $\ri\, \exp(-\ri\, \phi) = 1$, i.e., $\phi=\pi/2$.

\section{Details of the steps required for the quantum phase-gate operation}
In this appendix we consider the interaction of the atoms with the cavity vacuum and show how quantum phase gate operation could be achieved through this interaction.
For the purpose of this discussion we limit ourselves to only  the levels nearly resonant with the cavity, namely, $\ket{e_i}$ the excites state and $\ket{g_i}$ the ground state for  the $i$th atom in the cavity

The interaction the two atoms with the cavity is governed by the Hamiltonian
\begin{align}
{\mathcal H} &= \hbar \nu_k \, a_k^{\dagger} a_k + \hbar \omega_{e_1} \ket{e_1}\bra{e_1} 
+ \hbar \omega_{e_2} \ket{e_2}\bra{e_2} \nonumber \\ &-\hbar g ( \ket{e_1}\bra{g_1} + \ket{e_2}\bra{g_2} ) a_k - \hbar g (\ket{g_1}\bra{e_1}+\ket{g_2}\bra{e_2}) a_k^{\dagger} \nonumber \\
\end{align}
Note that we have taken both the atoms to be exactly identical and all the energies are measured with respect to the ground state of the atoms. In the main body of the paper we have used $\Omega_c$ for the cavity-atom coupling strength as opposed to $g$ here.

Now we consider possible initial states for the two atoms and explore their dynamical evolution individually to look for possible conditions under which the system returns to its initial state after interacting with the cavity for some time $\tau$. 
\begin{itemize}
\item
 $\mathbf{\ket{I_1} = \ket{g_1}\ket{g_2}\ket{0_k}}$\\
It can be easily seen that ${\mathcal H}  \ket{I_1}=0$. Thus, this state does not evolve in time.
\item
$\mathbf{\ket{I_2} = \ket{e_1}\ket{g_2}\ket{0_k}}$\\
We can see that this state is coupled to the states $\ket{g_1}\ket{e_2}\ket{0_k}$ and $\ket{g_1}\ket{g_2}\ket{1_k}$
Thus we can consider a general state $\ket{\Psi_2} = a(t) \ket{e_1}\ket{g_2}\ket{0_k} + b(t) \ket{g_1}\ket{e_2}\ket{0_k} + c(t) \ket{g_1}\ket{g_2}\ket{1_k}$ with $a(t=0)=1$, and $b(t=0)=c(t=0)=0$ and study its dynamics.
With the Schr\"{o}dinger equation
\begin{equation}
\ri\, \hbar \dot{\Psi} =  {\mathcal H} \Psi
\end{equation}
we obtain the following set of equations
\begin{align}
\dot{a}(t) &=  \ri\, g \,c(t) - \ri\, \omega_{e} a(t) \nonumber \\
\dot{b}(t) &=  \ri\, g \,c(t) - \ri\, \omega_{e} b(t) \\
\dot{c}(t) &=  - \ri\, \nu_k \,c(t) + \ri\, g \,a(t) + \ri\, g \,b(t) \nonumber
\end{align}
In the rotated frame defined by $a(t) = \exp(-\ri\, \omega_e  t) \tilde{a}(t)$, 
$b(t) = \exp(-\ri\, \omega_e  t) \tilde{b}(t)$ and 
$c(t) = \exp(-\ri\, \omega_e  t) \tilde{c}(t)$ the time derivatives can be written as
\begin{align}
\dot{a}(t) &= -\ri\, \omega_e \exp(-\ri\, \omega_e t) \tilde{a}(t)  + \exp(-\ri\, \omega_e t) \dot{\tilde{a}}(t)   \nonumber \\
\dot{b}(t) &= -\ri\, \omega_e \exp(-\ri\, \omega_e t) \tilde{b}(t)  + \exp(-\ri\, \omega_e t) \dot{\tilde{b}}(t) \\
\dot{c}(t) &= -\ri\, \omega_e \exp(-\ri\, \omega_e t) \tilde{c}(t)  + \exp(-\ri\, \omega_e
t) \dot{\tilde{c}}(t) \nonumber\, .
\end{align}
Thus,  the new rate equations, dropping the $\tilde{\phantom{x}}$, take the form
\begin{align}
\dot{a}(t) &= \ri\, g\, {c}(t)  \nonumber \\
\dot{b}(t) &= \ri\, g\, {c}(t)\\
\dot{c}(t) &= \ri\, (\omega_e-\nu_k) {c}(t) + \ri\, g \,(a(t) + b(t))  \nonumber
\end{align}
Noting that $\Delta=\omega_e-\nu_k\gg g$ we can set $\dot{c}(t)=0$ to 
obtain the
steady state value 
\begin{equation}
c(t)= -\frac{g}{\Delta} (a(t) + b(t)).
\end{equation}
Substituting this result in the  other rate equations we obtain
\begin{align}
\dot{a}(t) &= -\ri\, \frac{g^2}{\Delta} (a(t) + b(t)) \\
\dot{b}(t) &= -\ri\, \frac{g^2}{\Delta} (a(t) + b(t)) 
\end{align}
Thus with the initial condition $a(0)=1$ we obtain the solution
\begin{align}
\label{Eq:TwoLevelSolution}
a(t)&=\exp{(-\ri\,\frac{g^2}{\Delta}t)}\,\cos{(\frac{g^2}{\Delta}t)} \nonumber \\
b(t)&=-i \exp{(-\ri\,\frac{g^2}{\Delta}t)}\,\sin{(\frac{g^2}{\Delta}t)}
\end{align}
Thus for $(g^2/\Delta) t=\pi$ we have only the initial state populated. That is,
the final state is $\exp{(-\ri\,\pi)}\cos{\pi}\ket{e_1}\ket{g_2}\ket{0_k}$.
However this is the state in the rotated frame, we should go back to the lab frame which takes the form:
$\exp{(-\ri\, \omega_e t)}\ket{e_1}\ket{g_2}\ket{0_k}$, i.e., $\exp{(-\ri\, \pi \omega_e \Delta /g^2)}\ket{e_1}\ket{g_2}\ket{0_k}$, since $t = \pi\Delta/g^2$.
\item
$\mathbf{\ket{I_3} = \ket{g_1}\ket{e_2}\ket{0_k}}$ \\
In this case the roles of $a(t)$
and $b(t)$ are reversed, thus, at a time $t$ satisfying $(g^2/\Delta) t=\pi$ we
once again obtain the initial state back, i.e.,  $\exp{(-\ri\,\pi)} \cos{\pi}\ket{g_1}\ket{e_2}\ket{0_k}$. And in the lab frame it becomes $\exp{(-\ri\, \pi \omega_e \Delta /g^2)}\ket{g_1}\ket{e_2}\ket{0_k}$.
\item
$\mathbf{\ket{I_4} = \ket{e_1}\ket{e_2}\ket{0_k}}$\\
Once again it can be seen that this state will be coupled to
$\ket{g_1}\ket{e_2}\ket{1_k}$, $\ket{e_1}\ket{g_2}\ket{1_k}$ and these states
can be coupled to the state $\ket{g_1}\ket{g_2}\ket{2_k}$
states through the Hamiltonian under consideration. Thus we can take
\begin{multline}
\ket{\Psi_4} = a(t) \ket{e_1}\ket{e_2}\ket{0_k} + b(t)
\ket{g_1}\ket{e_2}\ket{1_k}\nonumber\\ + c(t) \ket{e_1}\ket{g_2}\ket{1_k} 
+ d(t)\ket{g_1}\ket{g_2}\ket{2_k}.
\end{multline}
The equations of motion can be written as
\begin{multline}
\dot{a}(t) =  \ri\, g \,(b(t)+c(t)) - 2 \ri\, \omega_{e} a(t)  \\
\dot{b}(t) =  - \ri\, \nu_k \,b(t) + \ri\, g \,(a(t)+ d(t)) - \ri\, \omega_{e} b(t) \\
\dot{c}(t) =  -\ri\, \nu_k \,c(t) + \ri\, g \,(a(t) + d(t)) - \ri\, \omega_{e} c(t)  \\
\dot{d}(t) =  - 2 \ri\, \nu_k \,d(t) +\ri\,g \,(b(t) + c(t)) 
\end{multline}
Once again moving to the rotated frame and sticking to the same notation we
obtain
\begin{multline}
\dot{a}(t) =  \ri\,g \,(b(t)+c(t))  \\
\dot{b}(t) = \ri\, \Delta \,b(t) + \ri\, g \,(a(t)+ d(t))  \\
\dot{c}(t) =  \ri\, \Delta \,c(t) + \ri\, g \,(a(t) + d(t))  \\
\dot{d}(t) =  2 \ri\, \Delta \,d(t) + \ri\, g \,(b(t) + c(t)) 
\end{multline}
Now with the assumption that $\Delta\gg g$ we can  obtain at steady state
\begin{equation}
d(t) = -\frac{g}{2\Delta} (b(t) + c(t))
\end{equation}
By substituting the steady state value of $d(t)$ in the rate equations for $b(t)$ and $c(t)$
and solving them at steady state gives,
\begin{align}
b(t)&=-\frac{g \Delta}{\Delta^2-g^2} a(t) \\
c(t)&=-\frac{g \Delta}{\Delta^2-g^2} a(t)\,. 
\end{align}
Using these solution we obtain
Thus we obtain
\begin{equation}
\dot{a}(t) = - 2 \ri\, \frac{g^2 \Delta }{\Delta^2-g^2} a(t) 
\end{equation}
which has a solution
\begin{equation}
a(t)=\exp{(-2 \ri\, \frac{g^2 \Delta }{\Delta^2-g^2} t)}= \exp{(-2 \ri\, \frac{g^2}{\Delta} t)}
\end{equation}
Here the last term is obtained by ignoring $g^2$ compared to $\Delta^2$ in the denominator and
simplifying.
Thus at time $t$ given by $g^2 t/\Delta=\pi$
the final state is given by $\ket{e_1}\ket{e_2}\ket{0_k}$ without any phase
factor. Howver, after undoing the transformation to the rotated frame it becomes 
$\exp{(-\ri\, 2 \pi \omega_e \Delta/g^2)}\ket{e_1}\ket{e_2}\ket{0_k}$
\end{itemize}
Thus it is clear that having these states as direct qubit combinations would not give us the phases needed to construct the quantum phase gate. We consider the effect of cavity interactions on some special states, where the target qubit, i.e., the second atom has been shifted to a state $\ket{a}$ whenever it starts with the excited state $\ket{e}$. This gives us the following possibilities for the two atom states: (i) $\ket{g_1}\ket{g_2}$ (ii) $\ket{g_1}\ket{a_2}$ (iii) $\ket{e_1}\ket{g_2}$ and (iv) $\ket{e_1}\ket{a_2}$. Now we arrange the level $a$ such that it does not interact with the cavity. We have already seen the evolution of possibilities (i) and
(iii). It only remains to be seen how the states $\ket{\rm ii}\equiv\ket{g_1}\ket{a_2}\ket{0_k}$ and 
$\ket{\rm iv}\equiv\ket{e_1}\ket{a_2}\ket{0_k}$ evolve. It is easy to see that the state $\ket{\rm ii}$ does not evolve; however, the state $\ket{\rm iv}$ can be shown to acquire a phase factor 
$\exp{(-\ri\, \pi \omega_e \Delta/g^2)}\exp{(-\ri\, g^2 t/\Delta)}$,
which for $\eta=g^2 t/\Delta=\pi$ is $-1$. The extra phase factor 
$\exp{(-\ri\, \pi \omega_e \Delta/g^2)}$ can be eliminated trivially. 

Using this we propose our scheme for the two-qubit gates as discussed in the text.
We choose the qubit states to be $\ket{0}$ and $\ket{1}$ which are not coupled to the cavity mode. Only when the interaction with the cavity is required state-transfer pulses are employed, through hamiltonian in Eq.~\eqref{Eq:ExtHamiltonian} to arrive from the initial qubit state to one of the (i), (ii), (iii), or (iv) states discussed above. Then the cavity interaction gives appropriate phase factors to the appropriate two qubit states facilitating the quantum phase gate. Then the qubits are transferred back from the levels interacting with the cavity to the long-lived states $\ket{0}$ and $\ket{1}$. This justifies the five steps needed to perform two qubit operation as discussed in the text.

To summarize the results of the atom-cavity interaction we see that:
\begin{widetext}
\begin{align*}
\ket{g_1}\ket{g_2}\ket{0_k} &\rightarrow \ket{g_1}\ket{g_2}\ket{0_k}  \\ 
\ket{e_1}\ket{g_2}\ket{0_k}&\rightarrow \exp{(-\ri\, \eta t)}\cos{\eta t} \ket{e_1}\ket{g_2}\ket{0_k} - \ri\, \exp{(-\ri\,\eta t)}\sin{\eta t} \ket{g_1}\ket{e_2}\ket{0_k}\\
\ket{g_1}\ket{e_2}\ket{0_k}&\rightarrow \exp{(-\ri\, \eta t)}\cos{\eta t}
\ket{g_1}\ket{e_2}\ket{0_k} - \ri\, \exp{(-\ri\, \eta t)}\sin{\eta t}\ket{e_1}\ket{g_2}\ket{0_k}\\
\ket{e_1}\ket{e_2}\ket{0_k}&\rightarrow \exp{(-2 \ri\, \eta t)} \ket{e_1}\ket{e_2}\ket{0_k}\\
\ket{g_1}\ket{a_2}\ket{0_k}&\rightarrow \ket{g_1} \ket{a_2}\ket{0_k} \\
\ket{e_1}\ket{a_2}\ket{0_k}&\rightarrow \exp{(-\ri\, \eta t)} \ket{e_1}\ket{a_2}\ket{0_k}
\end{align*}
where $\eta=g^2/\Delta$. Thus, for $\eta t = \pi$ we obtain
\end{widetext}
\begin{align}
\ket{g_1}\ket{g_2}\ket{0_k} &\rightarrow \ket{g_1}\ket{g_2}\ket{0_k}  \nonumber \\ 
\ket{e_1}\ket{g_2}\ket{0_k}&\rightarrow \ket{e_1}\ket{g_2}\ket{0_k} \nonumber \\
\ket{g_1}\ket{e_2}\ket{0_k}&\rightarrow \ket{g_1}\ket{e_2}\ket{0_k} \nonumber \\
\ket{e_1}\ket{e_2}\ket{0_k}&\rightarrow \ket{e_1}\ket{e_2}\ket{0_k} \nonumber \\
\ket{g_1}\ket{a_2}\ket{0_k}&\rightarrow \ket{g_1} \ket{a_2}\ket{0_k} \nonumber \\
\ket{e_1}\ket{a_2}\ket{0_k}&\rightarrow (-1) \ket{e_1}\ket{a_2}\ket{0_k}
\end{align}

In the text the auxiliary level $\ket{a_2}$ is actually the qubit state $\ket{1}$ which is not coupled to the cavity.

This discussion also explains the results presented in Table I.

\section{Fidelity calculation}
In this appendix we briefly discuss the fidelity calculation for the proposed non-local quantum phase gate.

We notice that when the initial state  of the two qubits is $\ket{e_c,g_t}$ there could be decoherence during the time the atoms are interacting with the cavity. If level $\ket{e_c}$ decays spontaneously. As we have discussed in the text, this state can be taken to be a metastable state of the atom, then decoherence does not really become an issue. Nevertheless, we carry out the analysis to see the possible effect of atomic decay and cavity  decay on the gate fidelity.

\begin{figure}[ht]
\includegraphics[scale=0.55]{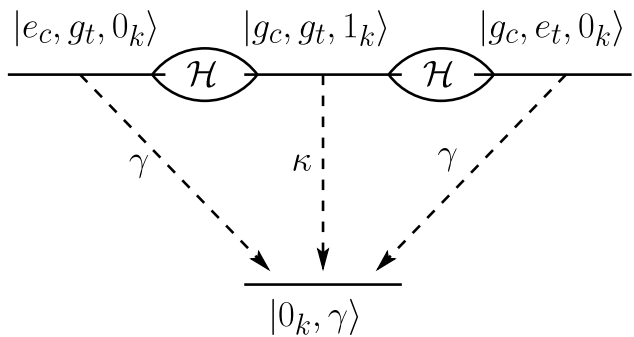}
\caption{\label{Fig:Decay}The decoherence mechanism for the two qubit gate during the interaction with the cavity through Hamiltonian ${\mathcal H}$. We note that the state after the spontaneous decay of the atom or the cavity decay is dynamically decoupled from the atom-cavity interaction Hamiltonian ${\mathcal H}$.}
\end{figure}

Here we would like to estimate the effect of decoherence caused by spontaneous emission from the level $\ket{e}$ to level $\ket{1}$. It can be noted that by clever choice of the quantum numbers for level $\ket{0}$ one can suppress spontaneous emission decay from level $\ket{e}$ to $\ket{0}$. Therefore we only need to consider spontaneous emission from level $\ket{e}$ to level $\ket{1}$. We should note that the initial state $\ket{e_c,g_t,0_k}$ couples to the states $\ket{g_c, e_t, 0_k}$ and $\ket{g_c, g_t, 1_k}$. Noting that spontaneous emission affects the level $\ket{e_t}$ and the cavity decay $\kappa$ affects the cavity state of $\ket{1_k}$ we need to include more states in our state space, namely, $\ket{1_c, g_t,0_k,\gamma}$, $\ket{g_c, 1_t,0_k,\gamma}$ and $\ket{g_c, g_t,0_k,\gamma}$. We term them collectively  as $\ket{0_k,\gamma}$.  Where  $\gamma$ corresponds to either the spontaneously emitted photon by the atoms or the photon decayed from the cavity. It is important to note that the state $\ket{0_k,\gamma}$ is dynamically decoupled from the rest of the states of interest.
The decay mechanism is illustrated in Fig.~\ref{Fig:Decay}
As it can be seen from Fig.~\ref{Fig:Decay} the state after the decay of the atoms of the cavity is dynamically decoupled from the atom-cavity interaction Hamiltonian, thus studying the dynamics of the amplitude equations is sufficient and complete density matrix treatment is not required. Let the general  state of the two atoms and cavity state  be
\begin{multline}
\ket{\Psi} = a(t) \ket{e_c, g_t, 0_k} + b(t) \ket{g_c, e_t, 0_k} \\
+ c(t) \ket{g_c, g_t, 1_k} + d(t) \ket{g_c, g_t, 0_k, \gamma}\,.
\end{multline}
The evolution of the state under the influence of the atom-cavity interaction Hamiltonian and the decay mechanisms gives the dynamical equations
\begin{align}
\dot{a}(t) &= - \frac{\gamma}{2} a(t) + \ri \,  \Omega_c c(t)\,,  \nonumber \\
\dot{b}(t) &=- \frac{\gamma}{2} b(t) + \ri\,  \Omega_c c(t)\,, \\
\dot{c}(t) &= (\ri\, \Delta - \frac{\kappa}{2} )c(t) + \ri \,  \Omega_c (a(t) +  b(t))\,. \nonumber 
\end{align}
Once again we adiabatically eliminate the state $\ket{g_c, g_t, 1_k}$ with the assumption that $\Delta \gg \Omega_c \gg \kappa$, to arrive at
\begin{equation}
c(t) = - \frac{\Omega_c}{ \Delta+\ri\,\kappa/2 } (a(t) +  b(t))\,.
\end{equation}  
Substituting $c(t)$ in the other equations we obtain
\begin{align}
\dot{a}(t) &= - \frac{\gamma}{2} a(t) - \ri\,  \frac{\Omega_c^2}{ \Delta+\ri\,\kappa/2 } (a(t) +  b(t))\,, \nonumber \\
\dot{b}(t) &=- \frac{\gamma}{2} b(t) - \ri\, \frac{\Omega_c^2}{ \Delta+\ri\,\kappa/2 } (a(t) +  b(t))\,.
\end{align}
The solution of the above equations with the initial condition $a(0)=1, b(0)=0$ is given by
\begin{widetext}
\begin{align}
a(t) &= \frac{1}{2} \exp\(-\frac{t \gamma \Delta }{2 \Delta + \ri\, \kappa}\) \[ \exp\( -\ri\, \frac{ t \gamma \kappa}{2 (2\Delta + \ri\, \kappa)}\) + \exp\(-\ri \, \frac{t(\gamma \kappa + 8 \Omega^2)}{2 (2\Delta + \ri\, \kappa)}\)\]\,, \nonumber \\
b(t) & =- \frac{1}{2} \exp\(-\frac{t \gamma \Delta }{2 \Delta + \ri\, \kappa}\) \[ \exp\(- \ri\, \frac{ t \gamma \kappa}{2 (2\Delta + \ri\, \kappa)}\) - \exp\(-\ri \, \frac{t(\gamma \kappa + 8 \Omega^2)}{2 (2\Delta + \ri\, \kappa)}\)\] \,.
\end{align}
\end{widetext}
Now we choose several values for the parameters and determine the fidelity 
\begin{equation}
F= |a(t)|^2,
\end{equation}
the results for different values of $\gamma$ and $\kappa$ measured in the units of $\Omega_c$ are summarized in the Table below:
\begin{table}[ht]
\caption{Fidelity for various system parameters at time $t = \pi/ \eta = \Delta \pi/ \Omega_c^2 $. All parameters are given in the units of $\Omega_c$. Common parameters are: $\Delta = 10\, \Omega_c$. We calculate the fidelity with the adiabatic elimination analytical results and through a complete numerical procedure without the adiabatic elimination of the $\ket{1_k}$ state of the cavity field.}
\begin{tabular}{|c|c|c|c|}
\hline
$\gamma$ & $\kappa$ & $F$ & $F'$ (Complete Numerical) \\
\hline
$0.001$ & $ 0.1$ & 0.939334 &  0.924248 \\
\hline
$0.01$ & $ 0.1$ & 0.707988 & 0.698862  \\
\hline
$0.1$ & $ 0.1$ & 0.0418878 &  0.0430447 \\
\hline
$0.1$ & $ 0.01$ & 0.0430785 &  0.0466712 \\
\hline
$0.01$ & $ 0.01$ & 0.728113 &  0.727542 \\
\hline
$0.001$ & $ 0.01$ & 0.966035 & 0.960429 \\
\hline
$0.001$ & $ 0.001$ & 0.968768 & 0.965277  \\
\hline
\end{tabular}
\end{table}


\end{document}